\documentclass[aps,prb,twocolumn,showpacs,refautoscript,reprint,longbibliography]{revtex4-2}
\usepackage{boldline,multirow}
\usepackage{float}
\usepackage{graphicx,natbib}
\usepackage{bm,amsmath,amssymb,wasysym,amsfonts,dcolumn}
\usepackage[colorlinks=true, urlcolor=blue, linkcolor=blue, citecolor=blue, pdfborder={0 0 0}]{hyperref}
\usepackage[usenames,dvipsnames]{xcolor}
\usepackage{lipsum}
\usepackage{epstopdf}
\usepackage{cleveref}
\crefname{figure}{Fig.}{Figs}
\crefname{table}{Table}{Tables}
\crefname{section}{Sec.}{Sections}

\newcommand{\ket}[1]{\left| #1 \right>} 
\newcommand{\bra}[1]{\left< #1 \right|} 

\newcommand{\rom}[1]{\mathrm{#1}}

\newcommand{\lsf}{l_{\rom{sf}}}
\newcommand{\lof}{l_{\rom{of}}}

\begin{document}

\title{Injection of orbital angular momentum into transition metals from first-principles}
		
\author{Max Rang}
\email[Email: ]{maxsrang@gmail.com}
\author{Paul J. Kelly\thanks{corresponding author}}
\email[Corresponding.author: ]{P.J.Kelly@utwente.nl}
\affiliation{Faculty of Science and Technology and MESA$^+$ Institute for Nanotechnology, University of Twente, P.O. Box 217,
	7500 AE Enschede, The Netherlands}

\date{\today}

\begin{abstract}
We use quantum mechanical scattering calculations implemented in a basis of tight-binding muffin-tin orbitals to calculate nonequilibrium spin and orbital currents in transition metals with a view to understanding the length scale on which they decay. 
In the case of spin currents, the relaxation length, called the spin-flip diffusion length, is reasonably well understood. 
We apply our experience with spin currents to study orbitally-polarized currents and find that they behave qualitatively differently. 
Upon injection from a lead, orbital currents decay within a few atomic layers contradicting the current interpretation of experimental results which appear to show exponential decay on the length scale of the spin-flip diffusion length and longer.
When spin-orbit coupling is included, the injected orbital current is partially converted into a spin current within a few atomic layers. 
This insight provides a new perspective on the physics of the orbital Hall effect.
\end{abstract}


\maketitle

\section{Introduction}

An orbital current, a current of orbital angular momentum (OAM) generated by the orbital Hall effect (OHE), is predicted to behave similarly to a spin current, a current of spin angular momentum (SAM) generated by the spin Hall effect (SHE) \cite{Sala:prr22, Fukami:natp25}. 
According to the conventional drift-diffusion model, spin currents decay exponentially on a length scale determined by the spin-flip diffusion length $\lsf$ \cite{vanSon:prl87, Valet:prb93}. 
By analogy, one may define the orbital diffusion length $\lof$ to be the corresponding length scale for orbital currents \cite{Sala:prr22}.
In recent experiments, $\lof$ has been inferred to be of order 50--60$\, \pm 15 \,$nm \cite{Choi:nat23} or $47\pm 11\,$nm \cite{Hayashi:cmp23} for Ti, 
$6.6 \pm 0.6 \,$nm \cite{Lyalin:prl23} or $6.1 \pm 1.7 \,$nm \cite{Lee:cmp21} for Cr and $68\pm 16\,$nm \cite{Hayashi:cmp23} or $\sim 80\,$nm \cite{Seifert:natn23} for $\alpha$-W. 
These lengths are much longer than the values of $\lsf$ reported for these systems, $13.3 \,$nm for Ti \cite{Du:prb14} and of order $2 \,$nm for $\alpha$-W \cite{Wang:prm18}; there appears to be a large discrepancy between the experimental value of $\lsf$ in $\alpha$-W and the value computed from first-principles scattering calculations, which put the value of $\lsf$ at $29.6\,$nm \cite{Nair:prl21}. 
However, in view of the large spread of $\lsf$ values extracted from experiments on Pt, ranging from 1 nm to 10 nm (see Table V in Ref.~\cite{Wesselink:prb19}), it would be prudent to wait for confirmation of the reported values of both $\lsf$ and $\lof$ for $\alpha$-W. 
The orbital diffusion length of $6.6 \pm 0.6\,$nm extracted from recent MOKE experiments on Cr \cite{Lyalin:prl23} is also comparable in size to a reported value of its spin-flip diffusion length, $l_{\rm sf}^{\rm Cr} \sim 4.5\,$nm \cite{Bass:jpcm07, Zambano:jmmm02}. 

In this manuscript we will use first-principles scattering calculations to generate currents of (OAM) in semiinfinite leads and inject them as nonequilibrium currents into a finite scattering region containing as many as 100,000 atoms. 
Our aim is to to determine the length scale on which they decay for a number of materials of current interest with a view to gaining insight at the atomic level so that experimental results become more easily interpretable. 
To achieve this we will extend a computational scheme previously developed \cite{Starikov:prb18, Wesselink:prb19} to calculate $l_{\rm sf}$ by studying the decay of a nonequilibrium current of SAM injected into a variety of materials.

\section{Methods}

\begin{figure}[b]
\includegraphics[width=8.6cm]{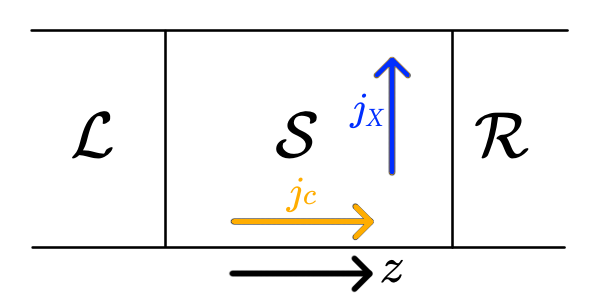}
\caption{Schematic of the left-lead$|$scattering-region$|$right-lead ($\mathcal{L}|\mathcal{S} |\mathcal{R}$) geometry. 
As a result of the spin/orbital Hall effects (SHE/OHE), a charge current $j_c$ originating in the leads parallel to the $z$ direction gives rise to a current $j_X$ of spin angular momentum (SAM, $X=s$) or orbital angular momentum (OAM, $X=l$) perpendicular to it.
Alternatively, by including a magnetic field in the left lead that couples to either the SAM or the OAM, $j_c$ can be polarized  directly.
Adapted with permission from Rang and Kelly \cite{Rang:prb25}. Copyright (2025) by the American Physical Society.
}
\label{fig1:schematic}
\end{figure}

The formalism used in the present work is Ando's Wave Function Matching (WFM) scheme \cite{Ando:prb91} implemented \cite{Xia:prb06, Starikov:prb18} in a basis of tight-binding muffin-tin orbitals (TB-MTO) \cite{Andersen:prl84, Andersen:85, Andersen:prb86}. 
The system under study comprises a scattering ($\mathcal{S}$) region sandwiched between semi-infinite leads on the left ($\mathcal{L}$) and on the right ($\mathcal{R}$), see \Cref{fig1:schematic}. 
The periodic boundary conditions in the transport direction of the crystalline leads are used to construct left- and right-propagating Bloch states which are incorporated into the Hamiltonian of the scattering region as self-energies \cite{Khomyakov:prb05}. 
In the end, a linear system of equations is solved to find the wave functions inside the scattering region. 
From these wave functions, one can extract intra-atomic properties like charge, spin and orbital densities, as well as inter-atomic properties like charge, spin and orbital currents \cite{Wesselink:prb19} induced by chemical potential differences in the linear response regime.
In studies where a current of SAM $j_{sz}^z$ (in the $z$ direction and polarized along $z$) was injected from a ferromagnetic lead material (or from a material made to be artifically ferromagnetic), the computational scheme proved efficient enough to allow calculations with thousands of atoms so that e.g. the interfaces between materials with different lattice constants could be modelled and the parameters governing their transport properties determined as a function of the temperature \cite{Gupta:prl20, Gupta:prb21, LiuRX:prb22, Gupta:prb22}.
More generally, we will be considering currents $j^{\beta}_{X\gamma}$ of quantity $X$ flowing in the direction $\beta$; when $X$ is SAM or OAM, the polarization direction is indicated by $\gamma$. 

Because we want to extend this scheme to currents of OAM, we briefly summarize it here.
Setting up one of the leads to be spin-polarized is straightforward. 
A constant spin-dependent potential is added to one of the spin channels so that all states of that spin are removed from the Fermi level. 
The corresponding term in the Hamiltonian describing this spin Zeeman interaction is
\begin{equation}
\label{Eq:sZ}
  \hat{V}_{\rm sZ} = B \left({\sigma}_z + \mathbb{I}\right)/2,
\end{equation}
where $\mathbb{I}$ is the identity matrix, $\sigma_z$ is the third Pauli spin matrix and $B$ is the magnitude of a magnetic field applied in the $z$-direction. 
The identity matrix needs to be included to shift the down-spin band back up, so that the original Fermi surface is restored \cite{Wesselink:prb19}. 

Although spin-orbit coupling (SOC) mixes up- and down-spin states, the newly constructed single spin Fermi surface is as similar as possible to the unperturbed one. 
An injected spin current decays exponentially on a length scale determined by the spin-flip diffusion length $\lsf$ \cite{Nair:prl21}; for Pt, an oscillation in the spin current immediately after the interface has been attributed to Fermi surface nesting and $\lsf$ is independent of the choice of electrode used for spin injection \cite{Wesselink:prb19}. 
SOC also leads to spin-polarized states acquiring a small orbital polarization so the charge current we inject from a spin-polarized lead also obtains an orbital polarization. 
When the SOC is switched off inside the scattering region, the spin polarization does not decay and we can study the orbital current on its own, assuming that the orbital diffusion length is independent of SOC which is not necessarily realistic.
Nevertheless, the qualitative behaviour of the orbital current should remain unaffected, especially in 3$d$ transition metals where the SOC is weak.

\subsubsection*{Kubo Formalism}
\label{sssec:Kubo}

The Kubo expression that is conventionally \cite{Tanaka:prb08, Go:prl18, Jo:prb18, Salemi:prm22} employed to calculate the ``$X$'' conductivity tensor, where $X$ is charge ($c$), spin angular momentum ($s$) or OAM ($l$), is
\begin{subequations}
\label{Eq:Kubo}
\begin{equation}
\sigma^{X\gamma}_{\alpha \beta} = \frac{e}{\hbar} \sum_n \int \frac{d^3{\bf k}}{\left(2\pi\right)^3} f_{n\mathbf{k}} \Omega^{X\gamma}_{n,\alpha \beta}({\bf k}),
\label{Eq:K1} \\
\end{equation}
where
\begin{equation}
\Omega^{X{\gamma}}_{n,\alpha \beta}({\bf k}) = 2\hbar^2 \!\! \sum_{m\ne n} \! {\rm Im} \! 
\left[  \frac{\bra{u_{n{\bf k}}} j^{\beta}_{X{\gamma}} \ket{u_{m{\bf k}}}    
        \bra{u_{m{\bf k}}}v_{\alpha}\ket{u_{n{\bf k}}}}  
        {\left(\varepsilon_{n{\bf k}} - \varepsilon_{m{\bf k}}+i \eta\right)^2} \right].
\label{Eq:K2}        
\end{equation}
\end{subequations}
$f_{n\mathbf{k}}$ is the Fermi-Dirac distribution, $\ket{u_{n\mathbf{k}}}$ is the cell-periodic part of the Bloch state with energy eigenvalue $\varepsilon_{n\mathbf{k}}$ and the velocity operator $\upsilon_\alpha$ is usually chosen to be  
\begin{equation}
\label{Eq:K3}
\upsilon_{\alpha} = \frac{1}{\hbar}\frac{\partial H({\bf k})}{\partial k_\alpha}
\end{equation}
in the crystal momentum representation \cite{Callaway:74}. 
In the limit that the inverse lifetime $\eta \sim 1/\tau \rightarrow 0$, $\Omega({\bf k})$ represents the Berry curvature and \eqref{Eq:K2} yields the conductivity but in practice finite values of $\eta$ are chosen to make convergence of the integral \eqref{Eq:K1} possible.
Performing the calculations in the primitive unit cell of the periodic crystal precludes the direct simulation of disorder.

In the scattering formalism, a spin current will not decay in the absence of disorder. 
Indeed, for an ideal bulk material, there will be no SHE either since states injected from the leads cannot scatter in the scattering region. 
The incident current is composed of all ``right-propagating'' states at the Fermi energy (for free electrons, the right hand side of the Fermi surface with positive velocities) without an explicit external electric field.
Assuming that the leads and scattering region are composed of the same materials, then in the absence of disorder and an external electric field, the incoming waves (labelled ${\bf k}_{\parallel}$) are solutions of the Schr\"odinger equation in the scattering region and just ballistically couple to the corresponding states in the right lead, with perfect transmission. 
At first glance, this might seem to contradict the conventional Kubo formula which is realistically only ever evaluated without disorder. 
In reality there is no discrepancy: the Kubo formula computes a spin Hall conductivity, not a current, and without a finite potential difference, there is no current. 
Alternatively, if the system is a perfect lattice, the electrical conductivity \eqref{Eq:K1} diverges in the limit $\eta \rightarrow 0$. 
Expressing the spin Hall angle (SHA) $\Theta_{\rm sH}$, which is what is calculated directly in the scattering formalism, as the ratio between the spin Hall conductivity $\sigma_{\rm sH}$ and the electrical conductivity $\sigma_c \equiv \sigma^c$
\begin{equation}
\Theta_{\rm sH} = \frac{\sigma_{\rm sH}} {\sigma_{\rm c}},
\label{EqII2}
\end{equation}
we directly see that in the perfect crystal limit the SHA is zero.
Our calculations are consistent with this: without any disorder in the scattering region, there is no lateral spin current.
Introducing any lattice disorder, no matter how small, results in a finite spin current. 

By analogy with \eqref{Eq:sZ}, we can define an orbital Zeeman term to add to the Hamiltonian of the left lead to realize orbital polarization. 
We introduce the potential
\begin{equation}
\label{Eq:oZ}
  \hat{V}_{\rm oZ} = B \ell_z ,
\end{equation}
where $\ell_z$ is the orbital angular momentum operator in the $z$ direction. 
This potential splits the bands though, as we will see, very high values of $B$ are needed to produce significantly polarized bands. 
The problem arises from the coupling of angular momentum and wave vector, which is much stronger than the SOC. 
Hence, constructing a Fermi surface that is similar to the equilibrium Fermi surface is challenging and interface effects are unavoidable. 
The weaker the coupling \eqref{Eq:oZ}, the less the left-lead bands are affected so the mismatch of states on either side of the interface is smaller; the downside is that the maximum orbital polarization achievable is correspondingly reduced.

The scattering problem involves solving the linear system of equations
\begin{equation}
\left(E_F\mathbb{I} - \hat{H}_S - \hat{\Sigma}_L - \hat{\Sigma}_R \right) \psi = Q_0,
\label{EqII4}
\end{equation}
where $\hat{H}_S$ is the Hamiltonian in the scattering region, $\hat{\Sigma}_{L/R}$ is the self-energy associated with the coupling of the scattering region to the leads, $E_F$ is the Fermi energy, $\mathbb{I}$ is the identity matrix, $\psi$ is the wave function we are looking for and $Q_0$ is a source term in the form of an incoming wave function, where nonzero elements are either right-propagating states in the left lead or left-propagating states in the right lead depending on whether we wish to calculate transmission probability amplitudes from left to right or from right to left; construction of the full scattering matrix requires both.  
We can add the orbital Zeeman term \eqref{Eq:oZ} to one or both leads, so only the self-energies $\Sigma$ and the source term $Q_0$ are affected by the additional potential $\hat{V}_{\rm sZ}$ or $\hat{V}_{\rm oZ}$. 
Alternatively, we can introduce a buffer zone inside the scattering region which is also polarized, in which case $\hat{H}_S$ also obtains the additional Zeeman term. 
The reason for doing this is that in the {\sc tqt} (Twente Quantum Transport) code \cite{TQT:GitHub} interatomic currents are only calculated inside the scattering region so introducing a few buffer layers allows us to image the currents just before the interface and gain additional insight.

\subsubsection*{Thermal lattice disorder}
\label{sssec:TLD}

To simulate a material at room temperature, we introduce thermal lattice disorder in the scattering region by displacing atoms from their equilibrium positions to construct snapshots of atoms vibrating at finite temperatures in the adiabatic approximation \cite{LiuY:prb11, LiuY:prb15}.
This method has been used to calculate various finite temperature spintronic properties such as the magnetization damping \cite{LiuY:prb11, Starikov:prb18}; $\l_{\rm sf}$ and the spin Hall angle for transition metals \cite{WangL:prl16, Wesselink:prb19, Nair:prl21}; the spin-memory loss (SML) parameter $\delta$ that is the interface analogue of $\l_{\rm sf}$ \cite{Gupta:prl20, Gupta:prb21, LiuRX:prb22, Gupta:prb22}.
We have empirically determined the values of the root mean square displacement $\Delta$ required to reproduce the experimental room temperature (RT) resistivity in the following.

\subsubsection*{Real-space currents}
\label{sssec:RSC}

We have argued that a current should refer to the flux of a local quantity \cite{Rang:prb24, Rang:prb25}. 
Using a definition of the expectation value of $\hat{\ell}$ that is  natural to the TB-MTO basis \cite{Andersen:prl84, Andersen:85, Andersen:prb86} and the atomic spheres approximation (ASA) \cite{Andersen:prb75}\begin{equation}
\ell_{\alpha,R} = \bra{\Psi_R} \hat{\ell}_\alpha \ket{\Psi_R},
\label{EqII5}
\end{equation}
together with the continuity equation, we can express the time variation of the $\alpha$th component of OAM on atom~$R$ as the net flow of OAM into the sphere at $R$ and a local torque 
\begin{equation}
\partial_t \ell_{\alpha,R} = -\iint_{S_R} \mathbf{j}_{\ell\alpha} \cdot d\mathbf{S} + \tau_{\alpha,R},
\label{EqII6}
\end{equation}
where the latter reflects the non-conserved nature of the OAM in a solid, in complete analogy with the SAM case \cite{Wesselink:prb19}. 
Here the wave function is expanded in the localized TB-MTO basis $|i \equiv Rlm\sigma \rangle$ where $R$ is an atomic site index and $lm\sigma$ have their conventional meaning so we can write
\begin{equation}
\!\! \ket{\Psi} \! = \! \sum_i |i\rangle \langle i| \ket{\Psi} 
                \! = \! \sum_R  \sum_{i_R} |i_R \rangle \langle i_R| \ket{\Psi} 
                \! = \! \sum_R \widehat{R} \ket{\Psi} ,
\label{EqII7}
\end{equation}
where $R$ runs over the atoms in the scattering region, $i_R$ sums over the orbitals on site $R$ and
$|\Psi_R \rangle = \widehat{R} |\Psi \rangle $
 \cite{Wesselink:prb19, Nair:prb21a}. 
Using the time-dependent Schr\"odinger equation and a decomposition of the flux into contributions associated with each atom with non-zero hopping, the current of the $\alpha$th component of the OAM ${\bm \ell}$ between atoms $P$ and $Q$ can be expressed as
\begin{equation}
\!\! j^{PQ}_{\ell\alpha} \! = \! \frac{1}{i\hbar} \left[ \bra{\Psi_P} \hat{\ell}_\alpha H_{PQ} \ket{\Psi_Q} - \bra{\Psi_Q} H_{QP} \hat{\ell}_\alpha \ket{\Psi_P}\right].
\label{EqII8}
\end{equation}
The derivation is straightforwardly generalized to any operator diagonal in atom-space, i.e. any operator $\widehat{A}$ for which
\begin{equation}
\bra{\Psi_P} \hat{A} \ket{\Psi_Q} \propto \delta_{PQ},
\label{EqII9}
\end{equation}
so that the flux of $\widehat{A}$ between atoms $P$ and $Q$ can be written as
\begin{equation}
\!\! j^{PQ}_{A} \! = \!  \frac{1}{i\hbar} \left[ \bra{\Psi_P} \widehat{A} H_{PQ} \ket{\Psi_Q} - \bra{\Psi_Q} H_{QP} \widehat{A} \ket{\Psi_P}\right].
\label{EqII10}
\end{equation}
This generalized flux reduces to the charge current when $\widehat{A} = 1$, to the current of the $\alpha$th component of SAM and OAM when $\widehat{A} = \hat{s}_\alpha$ and $\widehat{A} = \hat{\ell}_{\alpha}$, respectively.

Our so-called atom-centered approximation (ACA) \cite{Pezo:prb22} definition of the expectation value of $A$ contrasts with the ${\bf k}$-space formulation developed by Thonhauser \cite{Thonhauser:prl05}.
Unlike LAPW calculations where the orbital moments are calculated in (small) nonoverlapping muffin tin spheres \cite{Salemi:prm22}, our (large) overlapping atomic spheres are space-filling so there is no interstitial region whose contribution to the OAM is not included.

By considering the flux of OAM as it arises from the continuity equation \cite{Wesselink:prb19}, we sidestep the issue of having to define an OAM current operator as
\begin{equation}
\hat{j}_{\ell}=\frac{1}{2}{\left(\hat{\ell}\hat{v} + \hat{v} \hat{\ell}\right)},
\label{EqII11}
\end{equation} 
which effectively arose as the orbital equivalent of the conventional SAM current operator \cite{Shi:prl06}. 
Though conventional, the spin current operator
\begin{equation}
\hat{j}_{s}=\frac{1}{2}{\left(\hat{s}\hat{v} + \hat{v} \hat{s}\right)}
\label{EqII12}
\end{equation}
is not considered to be the ``proper'' operator \cite{Marcelli:ahp21}. 
In the case of spin currents, {\color{black} we are not aware of any calculations that predict significant quantitative differences between the operators}, but for the orbital currents, this  might not be the case. 

A similar procedure of first defining the expectation value of an observable $X$ and then using the continuity equation to define a current of this quantity has been adopted in recent publications by Valet {\it et al.} \cite{Valet:prb25, Valet:prl25}. 
However, their decomposition of $X$ into intra- and interband contributions made possible by working in momentum space is not possible in our real-space scattering formalism and will lead to results that appear at first sight to be different.

\section{Computational details}

The workflow of these calculations comprises several steps. 
First, the bulk atomic sphere potential must be generated with density functional theory (DFT). 
We used the {\sc questaal} code \cite{Pashov:cpc20}, with the von Barth-Hedin LDA functional \cite{vonBarth:jpc72} on a $\mathbf{k}$-point grid of $21 \times 21 \times 21$ and the tight-binding linear muffin-tin orbital basis \cite{Andersen:prl84, Andersen:85, Andersen:prb86}. 
The Kohn-Sham effective potential \cite{Kohn:pr65} is extracted and used as the input for the atomic sphere potentials of the {\sc tqt} transport code \cite{Wesselink:prb19, Starikov:prb18}. 
In this code, the wave-function matching problem \cite{Ando:prb91} is solved using periodic boundary conditions in the $x$ and $y$ directions, leaving $z$ as the transport direction, see \Cref{fig1:schematic}.
The scattering problem must be solved on a ${\bf k}$-point grid that is somewhat denser than the DFT calculation and here we have used a $19 \times 19$ grid for a $5 \times 5$ supercell, which is equivalent to a $95\times95$ grid for a $1\times 1$ unit cell.
We model thermal disorder by randomly displacing the atoms, drawing from a Gaussian distribution the variance of which reproduces the room temperature resistivity \cite{Wesselink:prb19, Starikov:prb18}. 
We generated ten configurations of random disorder and averaged over them to achieve adequate statistics. 

\section{Results}
\label{ssec:SOHC}

We begin with a brief review of some results for the spin and orbital Hall conductivities in \Cref{ssec:SOHC}. 
In \Cref{ssec:OPSOC} we consider how a current of SAM injected into Pt is converted by SOC into a current of OAM and in \Cref{ssec:OPZ} we use the orbital Zeeman interaction \eqref{Eq:oZ} to inject currents of OAM into Pt, Cr and V.

\subsection{Spin and Orbital Hall conductivities} 
\label{ssec:SOHC}

\begin{figure}[t]
\centering
\includegraphics[width=8.6cm]{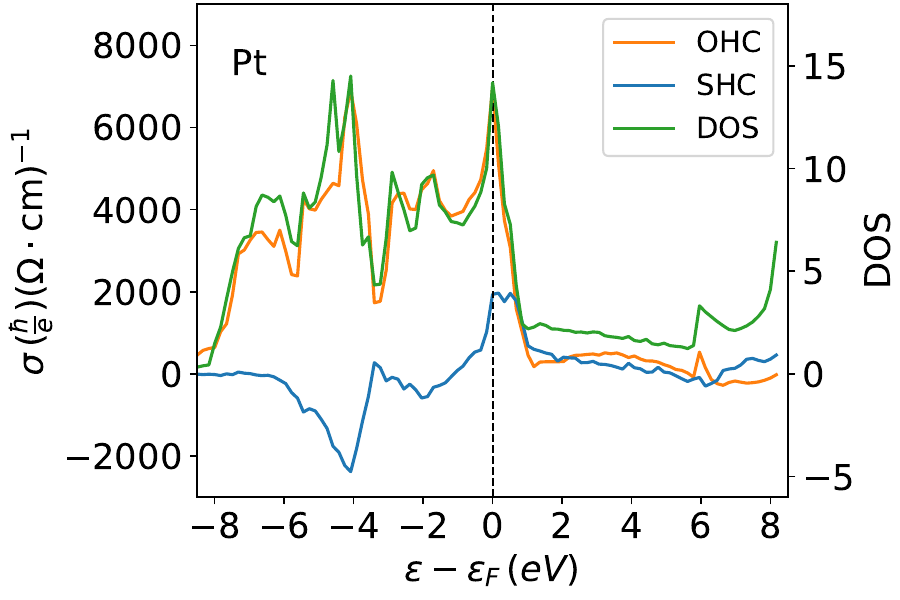}
\caption{Fermi energy dependence of (left-hand axis) the orbital (spin) Hall conductivity for bulk Pt at room temperature, meaning the atoms are randomly displaced from equilibrium by a Gaussian distribution with root mean square value $\Delta = 0.067 \AA$. The DOS is shown with respect to the right-hand axis. 
Reprinted with permission from Rang and Kelly \cite{Rang:prb25}. Copyright (2025) by the American Physical Society.}
\label{fig:eF_OHC_Pt}
\end{figure}

Our results \cite{Rang:prb25} for the SHC and OHC of fcc Pt are shown in \Cref{fig:eF_OHC_Pt}.
As a function of the band filling, the SHC $\sigma_{\rm sH}(\varepsilon)$ agrees reasonably well with results obtained \cite{Guo:prl08, Jo:prb18, Wesselink:prb19, Salemi:prm22} using the Kubo formalism \eqref{Eq:Kubo}, with a sharp peak at the Fermi level and another, negative, peak centered about 4~eV below the Fermi energy both of which will be slightly reduced when electronic temperature is introduced in the form of the Fermi-Dirac distribution. 
These peak structures can be related to the SOC-induced splitting of orbitally degenerate states at points and along lines of high symmetry \cite{Guo:prl08}. 
The OHC $\sigma_{\rm oH}(\varepsilon)$ is seen to have the same order of magnitude as $\sigma_{\rm sH}(\varepsilon)$, albeit larger, and its energy dependence maps the density of states $D(\varepsilon)$ very closely.
In spite of including thermal lattice broadening, we consistently find more structure in both  $\sigma_{\rm sH}(\varepsilon)$ and $\sigma_{\rm oH}(\varepsilon)$ with our scattering calculations than found using the Kubo formalism.
Part of this discrepancy may come from the huge value of lifetime broadening (0.4~eV) used by Salemi and Oppeneer \cite{Salemi:prm22} which, while perhaps appropriate for optical experiments, is less obviously justified for transport measurements where the lifetime diverges at the Fermi energy in the absence of disorder. 
Without knowing the value of the lifetime broadening used by Go {\it et al.} \cite{Go:prl18, Jo:prb18} we cannot speculate about the reason for the lack of structure in their $\sigma_{\rm oH}(\varepsilon)$. 

Focussing on the value of the OHC at the Fermi level, our scattering calculations predict a larger value of $\sigma_{\rm oH}(\varepsilon_F) \approx 7 \times 10^3 (\hbar / e) (\Omega \, \text{cm})^{-1}$ than the Kubo calculations. 
This large OHC value might partially explain the large variance in experimental values of the SHA in Pt (see Table V in \cite{Wesselink:prb19}). 
For example, in a Pt$|$FM bilayer geometry of Pt adjacent to a ferromagnet (FM), an orbital current generated in the Pt layer is injected simultaneously with a spin current, where the orbital-to-spin conversion at the interface and in the FM layer determines the magnitude of the resulting orbital torque \cite{Go:prr20a, Lee:cmp21, Lee:natc21}.
We expect that the interface conversion of orbital to spin current will depend on details of the interface atomic structure; this should be taken into account in the analysis of spin-orbit torque experiments in any case.

We used the same procedure to evaluate $\sigma_{\rm sH}(\varepsilon)$ and $\sigma_{\rm oH}(\varepsilon)$ for the elemental $3d$ metals Ti, V, Cr, and Cu, choosing values of $\Delta$  to reproduce the experimentally observed RT resistivities \cite{Rang:prb25}. 
In units of $10^3 (\hbar/e) (\Omega \, {\rm cm})^{-1}$, the values of $\sigma_{\rm oH}(\varepsilon=\varepsilon_F)$ we obtain for Ti, V, Cr and Cu are $5$, $6$, $2$ and $0.2$, respectively compared to the value of $7$ found for Pt.
For all four materials, the energy dependence of $\sigma_{\rm oH}(\varepsilon)$ tracks the density of states $D(\varepsilon)$ quite faithfully, in contrast to the results obtained with the Kubo equation \eqref{Eq:Kubo} \cite{Go:prl18, Jo:prb18, Salemi:prm22}. 
Because of the weakness of the SOC, $\sigma_{\rm sH}(\varepsilon_F) \sim 10 (\hbar/e) (\Omega \, {\rm cm})^{-1}$ for the 3$d$ metals.
The resistivity and orbital Hall angle $\Theta_{\rm oH} = \sigma_{\rm oH} / \sigma_{\rm c}$ are both found to be linear in temperature so $\sigma_{\rm oH}$ is at most weakly temperature dependent \cite{Rang:prb25}. 

\subsection{Orbital polarization through spin polarization}
\label{ssec:OPSOC}

\begin{figure}[t]
\centering
\includegraphics[width=8.4cm]{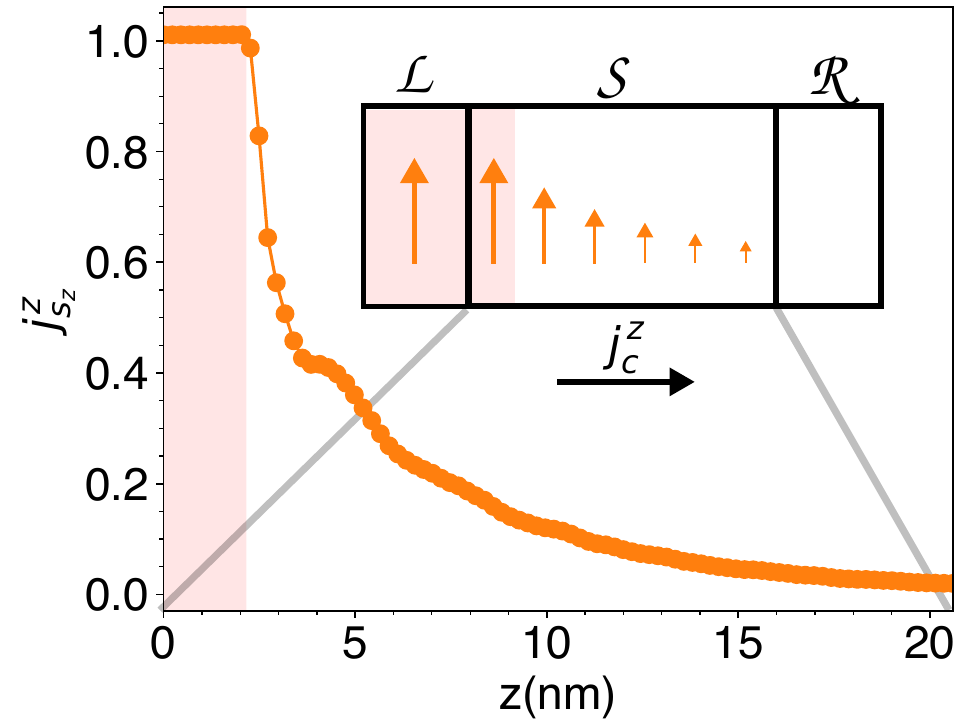}
\caption{Exponential decay of a spin-polarized current injected from a ballistic Pt lead into thermally disordered Pt. 
To visualize the current in the lead, the interface between ballistic (pink) and diffusive Pt is displaced into the scattering region to $z=z_I$ \cite{footnote2}. 
Reprinted with permission from Rang and Kelly \cite{Rang:prb24}. Copyright (2024) by the American Physical Society.}
\label{fig:lsf}
\end{figure}

In reference \cite{Wesselink:prb19}, an essentially fully spin-polarized charge current was constructed by choosing $B$ in \eqref{Eq:sZ} to be larger than the $d$ bandwidth.
On injecting this current of SAM into Pt, we could study the decay of the spin polarization and found it to be exponential over some five orders of magnitude (\Cref{fig:lsf}) allowing the spin-flip diffusion length $l_{\rm sf}$ to be determined.
The value of $l_{\rm Pt} \equiv l_{\rm sf}^{\rm Pt} \approx 5.2 \,$nm was found to be independent of the choice of material used in the left-hand lead and the procedure was extended to study $l_{\rm sf}$ for all 5$d$ elemental metals \cite{Nair:prl21} and to examine the SML parameter $\delta$ and its temperature dependence \cite{Gupta:prl20} for a number of important interfaces between two nonmagnetic (NM)   materials (NM$|$NM$'$) \cite{Gupta:prb22, LiuRX:prb22} and between nonmagnetic and  ferromagnetic (FM) materials (NM$|$FM) \cite{Gupta:prb21, LiuRX:prb22}.  

In a relativistic calculation, the SAM and OAM are coupled by the SOC and injection of a current of SAM is necessarily accompanied by a current of OAM \cite{Rang:prb24}. 
The magnitude of the orbital polarization depends on the strength of the SOC and for Pt the signal can be distinguished from the noise inherent to a numerical study of transport when disorder is included.
This will not be the case for the 3$d$ elements and in \Cref{ssec:OPZ} we will instead use \eqref{Eq:oZ} to produce a current of OAM.

\subsubsection{Platinum}

\begin{figure}[t]
\includegraphics[width=8.6cm]{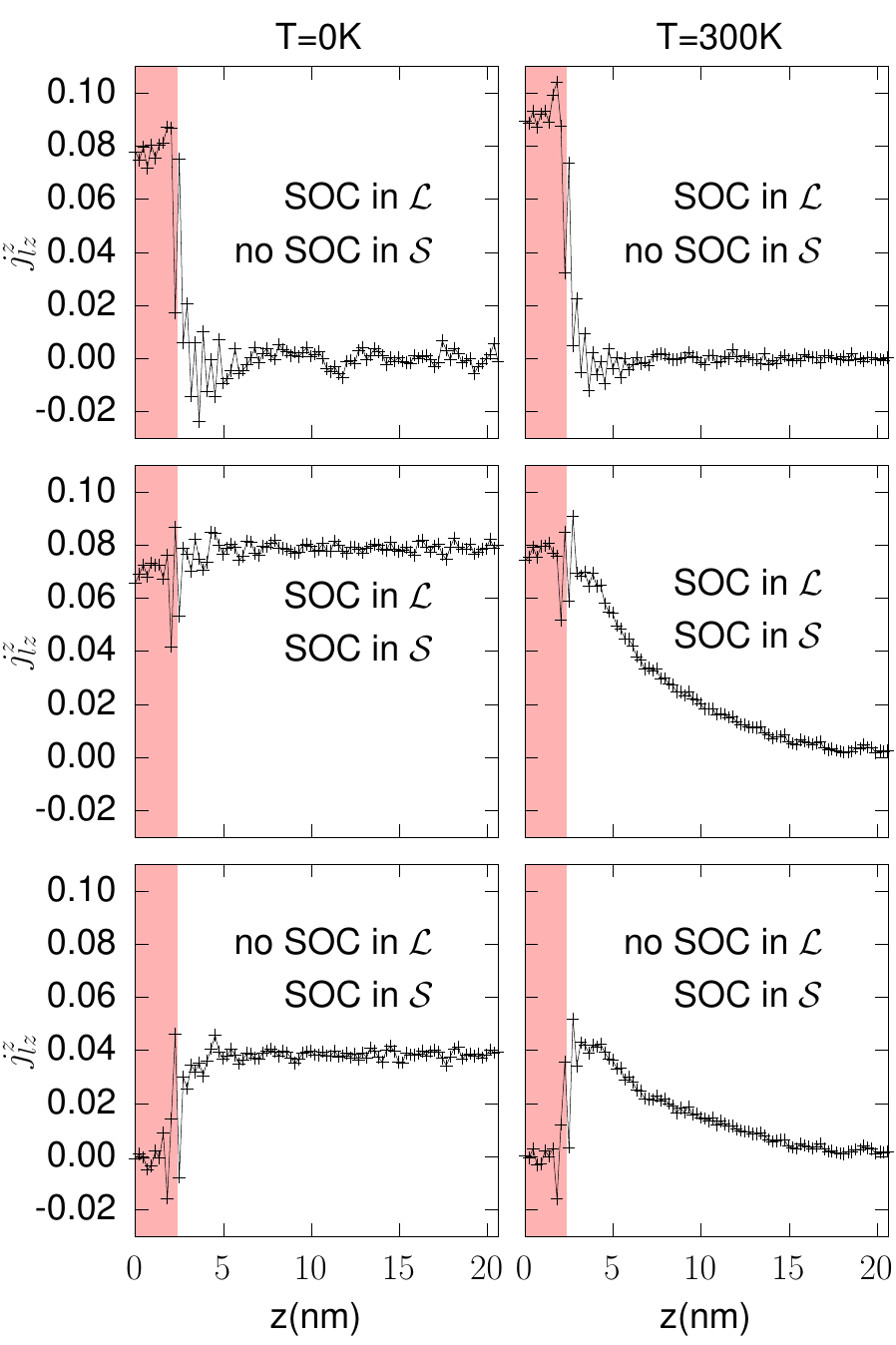}
\caption{Injection of a current of OAM polarized in the $z$ direction, $j_{lz}^z$, from a spin-polarized (in the $z$ direction) Pt lead (shaded rose) into Pt without (left column) and with (right column) thermal lattice disorder corresponding to $T=300\,$K. 
}
\label{fig:Pt_lcurs}
\end{figure}

With the highest spin Hall angle of any elemental metal, Pt is the default material for generating spin currents \cite{Hoffmann:ieeem13, Sinova:rmp15}. 
Although its spin-flip diffusion length $\lsf$ has been well studied, there is still no consensus about its value; the computational method we use here predicts $\lsf$ to be $\approx 5.2\,$nm for Pt at ``room temperature'', $T=300\,$K \cite{Wesselink:prb19}. 
Adding the spin-dependent potential described by \eqref{Eq:sZ} leads to the injection of currents of SAM and OAM into Pt whose decay can be studied to see how it depends on SOC (with or without) and thermal lattice disorder (with, $T=300\,$K and without, $T=0\,$K).
The resulting current distributions as a function of position $z$ inside the scattering region are shown in \Cref{fig:Pt_lcurs} for three systems: 
where the SOC is switched off inside the scattering region (top row);
where the SOC is switched on in the scattering region (middle row);
where the SOC is switched off in the left lead (bottom row). 
In the figure, the rose-shaded region is a buffer layer with the same properties as the lead introduced purely for visualization purposes; it is to all intents and purposes the left lead \cite{footnote2}. 

\begin{figure}[t]
\includegraphics[width=8.6cm]{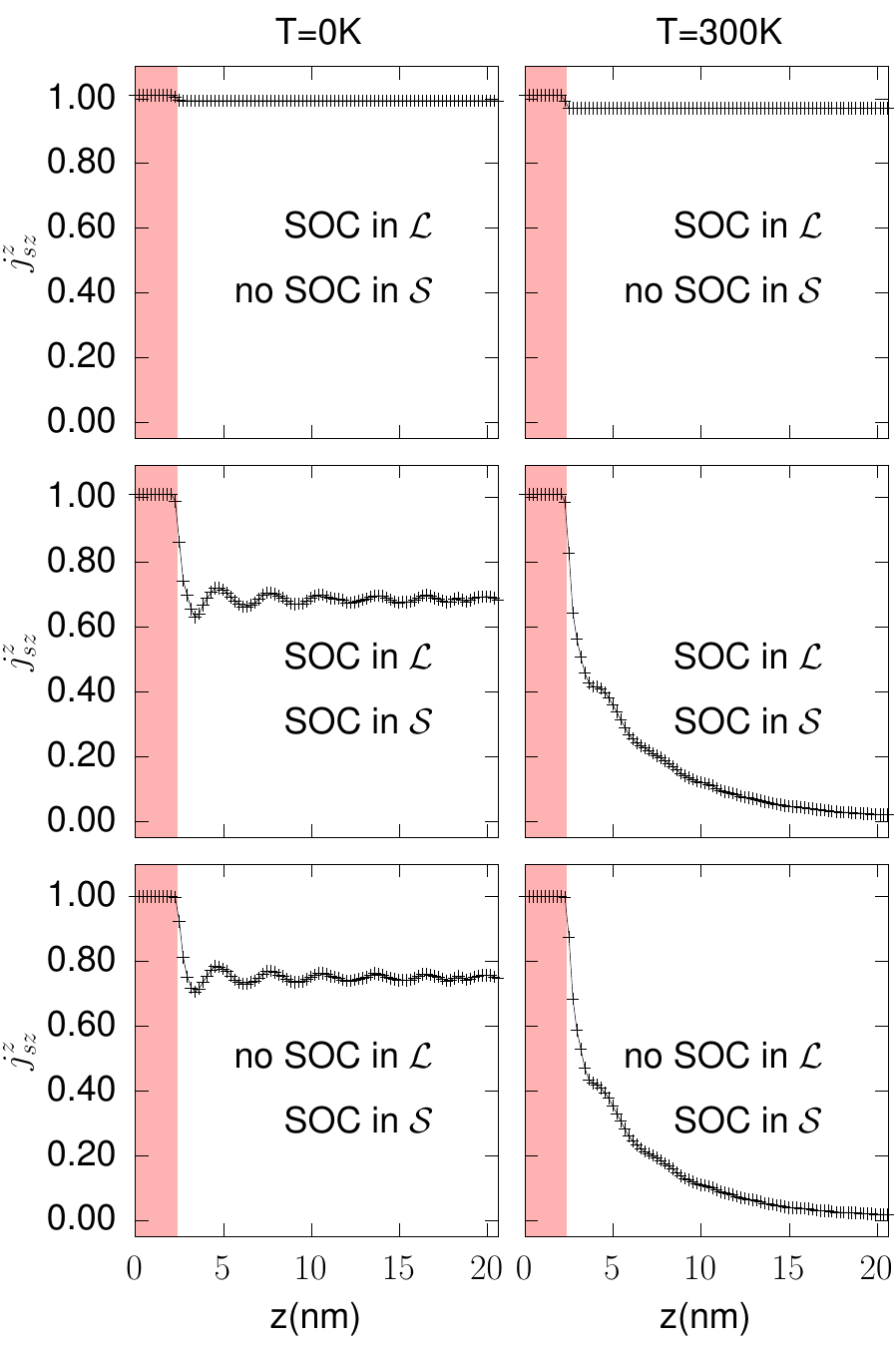}
\caption{Injection of a current of SAM polarized in the $z$ direction, $j_{sz}^z$, from a spin-polarized (in the $z$ direction) Pt lead (shaded rose) into Pt without (left column) and with (left column) thermal lattice disorder corresponding to $T=300\,$K.
}
\label{fig:Pt_scurs}
\end{figure}

In the first two cases, the SOC in the left lead is turned on so the incoming wave is orbitally polarized as well as spin polarized; the orbital polarization of the injected OAM current is seen to be approximately $0.08 \hbar$. 
In the top row with no SOC in the scattering region, the orbital polarization drops off very rapidly, independent of whether the Pt is disordered (right column) or not (left column). 
Switching the SOC on inside the scattering region changes this significantly (middle row).
Without disorder, the orbital polarization is more or less unaffected and stays constant at roughly the value it was initially injected with. 
With disorder, the orbital polarization decays exponentially with a decay rate roughly equal to the spin-flip diffusion length $\lsf \approx 5.2\,$nm.
Remembering that these orbital currents are coupled to an injected spin current, we can make sense of these observations by imaging the spin currents from which these orbital currents result in \Cref{fig:Pt_scurs}. 
The middle-row, right-column (middle-right; identical to \Cref{fig:lsf}) panel corresponds to precisely those conditions used previously to compute $\lsf$ \cite{Wesselink:prb19}. 
The orbital polarization in these calculations does not decay on its own; it is through its coupling to the spin degree of freedom that exponential decay occurs. 
In the absence of disorder (middle-left), a $z$-independent orbitally-polarized current (\Cref{fig:Pt_lcurs}) is induced by SOC to the essentially constant injected spin current (\Cref{fig:Pt_scurs}).
Without SOC, the orbital polarization decays within a few layers, as seen in the top row of \Cref{fig:Pt_lcurs}.

\begin{figure}[t]
\includegraphics[width=8.7cm]{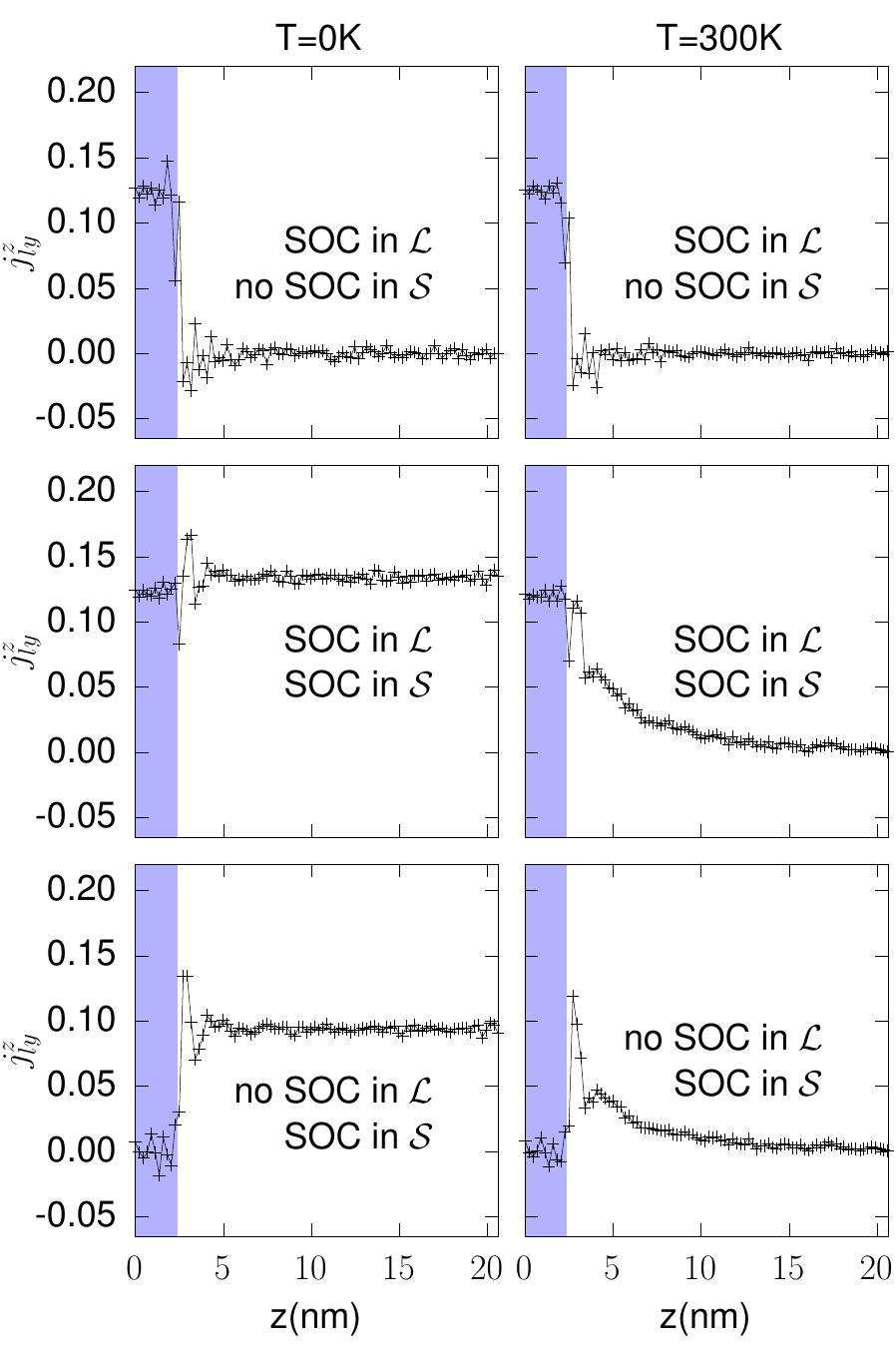}
\caption{Injection of a current of OAM polarized in the $y$ direction, $j_{ly}^z$, from a spin-polarized (in the $y$ direction) Pt lead (highlighted blue) into Pt without (left column) and with (right column) thermal lattice disorder corresponding to $T=300\,$K.
}
\label{fig:Pt_lycurs}
\end{figure}

The coupling of the orbital polarization to the spin polarization becomes even clearer in the bottom row of \Cref{fig:Pt_lcurs} where the SOC is turned off in the lead so there is no orbital polarization to be injected. 
Nevertheless, because SOC is switched on in the scattering region, the orbital current grows to a nonzero value that is, however, lower than when the injected current is already orbitally polarized (top row). 
In the absence of disorder (bottom-left), a constant orbitally polarized current is induced by the SOC to the constant spin polarized current.
With disorder, we again see the exponential decay only through the coupling to the spin, which shows a decay consistent with $\lsf \approx 5.2\,$nm. 
The length scale on which the spin current generates an orbital polarization in the bottom row at the no-SOC-in-$\mathcal{L}|$SOC-in-$\mathcal{S}$ interface is very comparable to what happens in the top row at the SOC-in-$\mathcal{L}|$no-SOC-in-$\mathcal{S}$ interface.
There, the absence of SOC in the scattering region forces the orbital polarization to decay rapidly.
In the bottom row, the orbital polarization of the injected current is initially zero because there is no  SOC in the lead and polarizes rapidly inside the scattering region because of the SOC there.

\begin{figure}[t]
\includegraphics[width=8.7cm]{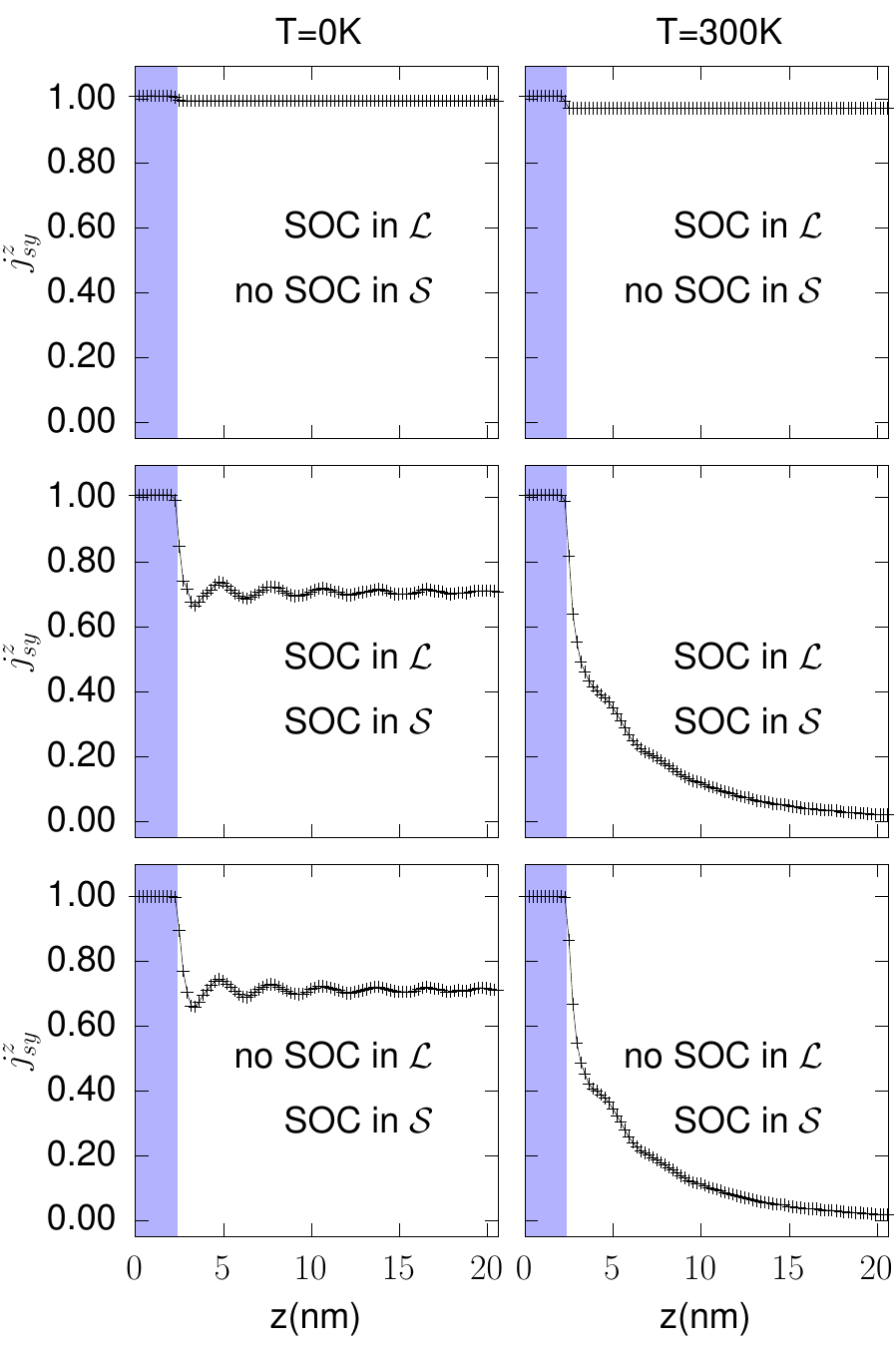}
\caption{Injection of a current of SAM polarized in the $y$ direction, $j_{sy}^z$, from a spin-polarized (in the $y$ direction) Pt lead (highlighted blue) into Pt without (left column) and with (right column) thermal lattice disorder corresponding to $T=300\,$K.
}
\label{fig:Pt_sycurs}
\end{figure}

The results discussed so far have included currents $j_{Xz}^z$ of SAM and OAM generated by a spin-polarized lead polarized in the $z$ direction which coincides with the charge current direction. 
In the case of the orbital (spin) Hall effect, the polarization of the orbital (spin) current is perpendicular to the charge current direction. 
To test if this makes a difference in our case, we rotate the quantization axis of the spin from the $z$ to the $y$ direction. 
The results are shown in \Cref{fig:Pt_lycurs} and \Cref{fig:Pt_sycurs}.
The currents of SAM in the latter are not altered at all; just as in the $z$-polarization case, exponential decay only occurs when the SOC is turned on and the lattice is disordered (middle-, bottom-right). 
For currents of OAM, the main difference between the $y$- and $z$-polarized cases is a slight peak right after the interface, most prominent in the bottom-right panel of \Cref{fig:Pt_lycurs}.

These results seem to disagree with the prediction of long range OAM currents through the notion of ``hot spots'' on the Fermi surface \cite{Go:prl23}, since we do not observe the predicted long range behaviour (save for the decay on the length scale of the spin flip diffusion length).
The orbital degeneracies upon which the ``hot spot'' mechanism relies should be present in our calculations. 
We conclude that they are either lifted by the thermal lattice disorder included in our calculations and/or the volume of phase space where they occur is too small for them to make a significant contribution.

\begin{figure}[t]
\includegraphics[width=8.6cm]{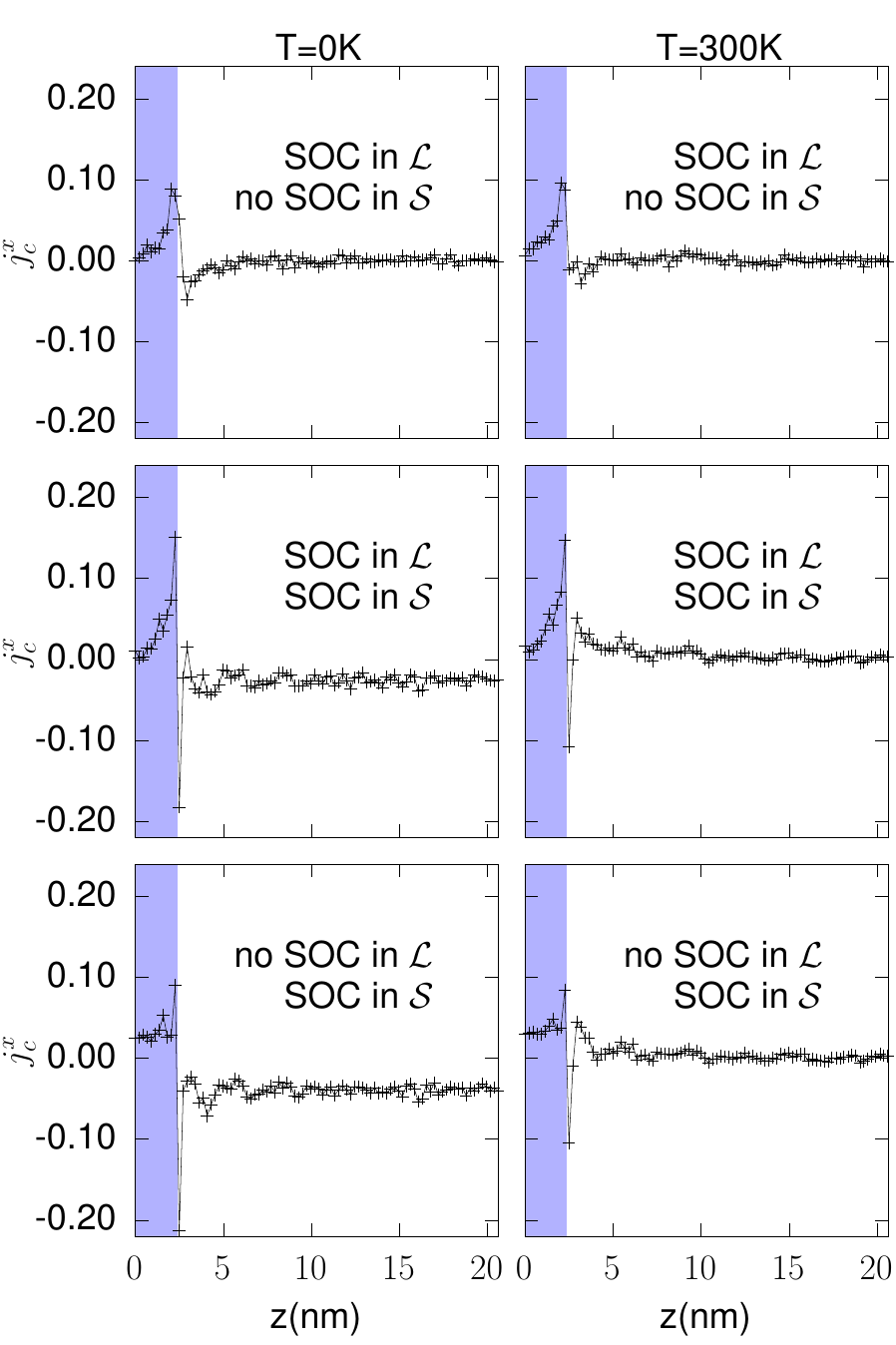}
\caption{Inverse spin Hall effect in Pt due to the spin current polarized in the $y$ direction injected from the left lead (highlighted in blue) into Pt without (left column) and with (right column) thermal lattice disorder corresponding to $T=300\,$K.
}
\label{fig:Pt_ccurs}
\end{figure}

Another feature we expect given the symmetry of the injected currents is an inverse spin Hall current. 
This charge current should flow in a direction perpendicular to the flow ($z$) and polarization ($y$) directions of the spin current, so we expect the inverse spin Hall current to be in the $x$ direction. 
This is precisely what we see in \Cref{fig:Pt_ccurs} though the effect is only visible when there is SOC in the scattering region (middle and bottom panels). 
In the presence of thermal lattice disorder in the scattering region (middle- and bottom-right), the spin current $j^z_{sy}$ decays exponentially on the length scale of $\l_{\rm sf}$ (\Cref{fig:Pt_sycurs}). 
In the scattering region, there are both interface and bulk contributions to the inverse spin Hall effect (ISHE) charge current $j^x_c$ \cite{WangL:prl16}.
Because of the weakness of the spin-charge conversion, $\Theta^{\rm Pt}_{\rm sH} \sim 4\%$ \cite{Wesselink:prb19}, $j^x_c$ is very small and becomes indistinguishable from noise on the length scale of $\l_{\rm sf}$ (\Cref{fig:Pt_ccurs}). 

In the absence of thermal disorder (middle- and bottom-left panels), the spin current drops at the interface to an essentially $z$-independent constant value (\Cref{fig:Pt_sycurs}) reflecting the divergence of $\l_{\rm sf}$ when $T=0\,$K.
The ISHE charge current $j^x_c$ is equal to the injected spin current times $\Theta^{\rm Pt}_{\rm sH}$ which means that if a fully spin-polarized current is injected, the resulting charge current can at most be equal to the amplitude of $\Theta^{\rm Pt}_{\rm sH}$ which is roughly $4\%$ in Pt \cite{Wesselink:prb19}. 
This is approximately the value we see in the middle- and bottom-left panels of \Cref{fig:Pt_ccurs}. 
Because there is no ISHE charge current in the absence of disorder, the finite values of $j^x_c$ seen in \Cref{fig:Pt_ccurs} are, perhaps counterintuitively, an interface effect; they result from (i) interface scattering and (ii) the divergence of $\l_{\rm sf}$.
The difference between the middle and bottom-left panels is a result of the presence or absence of SOC in the lead. 

\subsection{Orbital polarization using an orbital Zeeman interaction}
\label{ssec:OPZ}

By adding a site-diagonal orbital Zeeman interaction \eqref{Eq:oZ} coupling the OAM to a magnetic field of magnitude $B$ to the lead Hamiltonian, we can orbitally polarize the leads without directly involving the spin degree of freedom. 
Interestingly, nonzero spin-orbit coupling (SOC) will now lead to the converse of the effect seen before; orbital polarization of the leads necessarily introduces a small degree of spin polarization. 
We will see, however, that the maximum orbital polarization we can create with ``reasonable'' magnetic fields is quite small so that the resulting spin polarization is almost indistinguishable from noise, in particular when the SOC is very small.

\subsubsection{fcc Pt}

\begin{figure}[t]
\includegraphics[width=8.6cm]{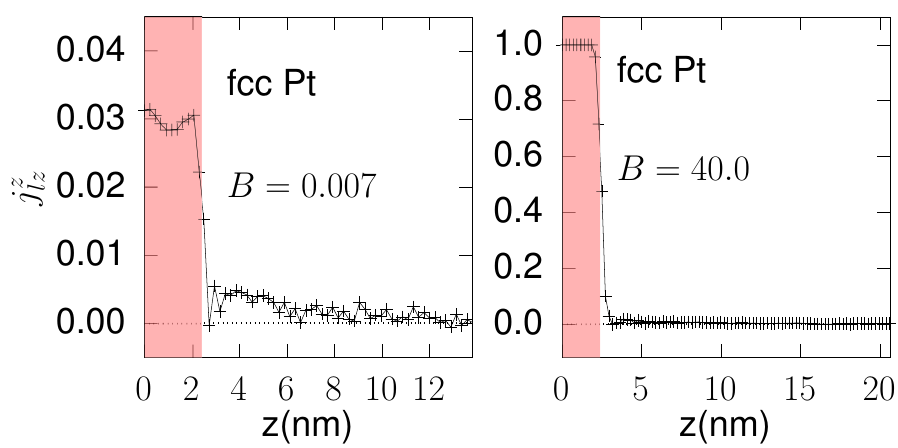}
\caption{Injection of a current of OAM, $j_{lz}^z$, into room temperature thermally disordered fcc Pt  from a lead with nonzero orbital magnetic fields $\hat{V}=B \ell_z$. 
The rose region is where the orbital Zeeman term is added to the Hamiltonian.
}
\label{fig:Pt_via_bl}
\end{figure}

We test the method outlined above by considering the orbital polarization that can be achieved in two extreme cases: 
(i) for a very large $B$-field with an amplitude $B \ell_z = 40$ Ryd and 
(ii) for a much smaller $B \ell_z = 7\,$mRydberg $\sim 0.1\,$eV. 
In the first case, we achieve total orbital polarization of a $p$ state $\langle L\rangle = \hbar$, though only when simultaneously tuning the Fermi energy to $26.8$ Ryd lower than normal. 
In the case of the small Zeeman field, we need to shift the Fermi energy up by $0.06$ Ryd to achieve the maximum polarization. 
\Cref{fig:Pt_via_bl} shows that in both cases the injected current of OAM loses almost all its orbital polarization within a few atomic layers. 
Note that the vertical scales in \Cref{fig:Pt_via_bl} are very different because the maximum orbital polarization attainable (by tuning the Fermi energy) with the low field is only $\approx 0.03 \hbar$.
The results for low polarization show that inside the Pt, the orbital current is positive for some nanometers inside the scattering region, which might be interpreted as an orbital diffusion. We recall the results presented earlier, where an injected spin current carries an associated orbital current. The same is happening here; the orbital current we inject is converted into a spin current within a few layers, after which the spin current itself generates an orbital current which is attenuated on a length scale of the spin-flip diffusion length.

\subsubsection{bcc Cr}

\begin{figure}[b]
\includegraphics[width=8.6cm]{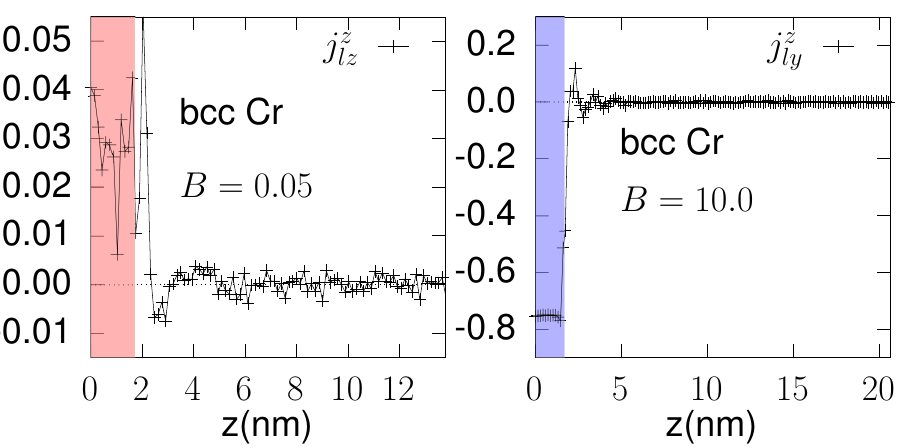}
\caption{Injection of an orbitally polarized current into room temperature nonmagnetic bcc Cr from a lead with a nonzero orbital magnetic field $\hat{V}=B \ell_z$. 
The region shaded rose is where the Zeeman-like term is added to the Hamiltonian, with the field oriented in the $z$-direction (left panel). The blue region is similar, but oriented in the $y$-direction (right panel).
}
\label{fig:Cr_via_bl}
\end{figure}

Theoretical predictions for Cr \cite{Jo:prb18, Salemi:prm22, Rang:prb25} identify it as a promising ``orbitronic'' material with a high orbital Hall conductivity. 
Recent experimental results appear to confirm this \cite{Lee:cmp21, Lyalin:prl23}. 
These experiments suggests an orbital diffusion length of $\lof \approx 6 \,$nm. 
By injecting an orbitally polarized current into bulk bcc Cr we can directly calculate the orbital current and its decay \cite{Rang:prb24}. 
In \Cref{fig:Cr_via_bl} we see that within a few atomic layers the injected orbital current is reduced to the level of the numerical noise associated with the thermal lattice disorder. 
The Fermi energy in the lead was shifted by $0.07$ Ryd to maximize the orbital polarization. 
Two different orientations of the ``magnetic field'' show the same short range of the injected orbital current.
Note that the relatively long-range orbital currents we saw in Pt are absent here because of the very small SOC (and correspondingly small spin Hall angle) in Cr so that the injected orbital current is converted into a negligible spin current.

\subsubsection{bcc V}
\begin{figure}[t]
\includegraphics[width=8.6cm]{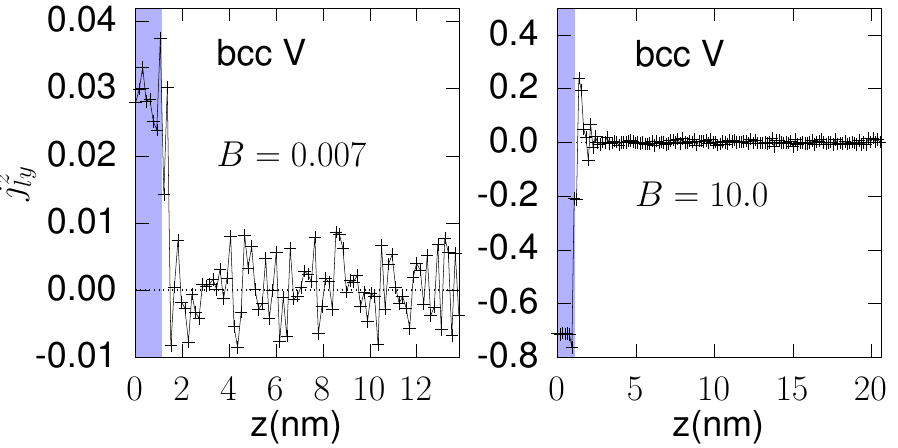}
\caption{Injection of an orbitally polarized current into room temperature bcc V from a lead with a nonzero orbital magnetic field $\hat{V}= B \ell_y$. The blue region is where the Zeeman-like term is added to the Hamiltonian.
}
\label{fig:V_via_bl}
\end{figure}

To confirm that the calculations for Cr hold for other 3$d$ transition metals, we repeat the above calculations for V \cite{Rang:prb24}. 
In \Cref{fig:V_via_bl}, we observe precisely the same behaviour, i.e., the injected orbital current is reduced to the numerical value of the noise intrinsic to the disordered calculations within a few atomic layers.
To show that the polarization direction of the orbital current makes no difference, we oriented the ``magnetic field'' in the $y$-direction, constructing the orientation of the orbital Hall effect{\color{black}, i.e. such that the polarization of the orbital moment is perpendicular to the current direction}.
This has no effect on the results, the orbital current vanishes within a few atomic layers.
These results are very similar to the Cr case except that the noise is larger when the polarizing field and polarization of the injected current is smaller.

\section{Discussion}
\label{sec:disc}

The experimental setup most commonly used to study the spin Hall and orbital Hall effects is a so-called current-in-plane (CIP) configuration whereby a charge current is passed through a NM$|$FM bilayer parallel to the interface leading to a transverse spin or orbital Hall current perpendicular to the interface.
The transverse current is ``pure'', unaccompanied by a charge current and occurs in nonmagnetic materials without breaking time-reversal symmetry. 
We have performed CIP calculations for NM$|$FM bilayers (unpublished) but just as for the original work on the giant magnetoresistance (GMR) effect \cite{Levy:ssp94, Gijs:ap97, Tsymbal:ssp01}, the CIP configuration is very complex and our results do not lend themselves to simple interpretation. 
For a CIP NM$|$FM bilayer, there are not only spin and orbital currents arising from the SHE and OHE in the NM layer but also spin and orbital currents in the FM as well as ``in'' the interface itself \cite{Hou:apl12, WangL:prl16, Amin:prb16a, *Amin:prb16b}, not to mention Rashba-Edelstein effects (to the extent that they are actually different effects). 
In addition, spin-orbital and orbital-spin interconversion occurs at the interfaces as well as the ``usual'' interface effects like spin-dependent transparency, spin-memory loss etc. most of which are not included in interpreting OHE experiments. 
Many experiments were carried out with NM$|$NM$'|$FM trilayers involving NM$'$ spacer layers with a corresponding increase in the number of unknown parameters. 

Because of this greater complexity, the study of GMR rapidly shifted to the current-perpendicular-to-the-plane (CPP) configuration because of its higher symmetry \cite{Levy:ssp94, Gijs:ap97, Tsymbal:ssp01, Bass:jmmm16}. 
In the context of spin transport, we are not aware of any suggestions that the parameters entering the Valet-Fert description of CPP-GMR \cite{Valet:prb93} should be invalid in the CIP case. 
Nor are we aware of suggestions that the pure spin currents encountered in spin-pumping or SHE experiments should be described by different parameters than those used to describe the behaviour of spin-polarized charge currents. 
When we study the behaviour of an orbitally polarized current prepared by adding the time-reversal symmetry breaking term \eqref{Eq:oZ} to the Hamiltonian for the left lead, we make the tacit assumption that currents of OAM, like pure spin (SAM) currents, can be decomposed into cancelling charge currents with opposite angular momenta that can be studied individually and that the parameters describing their behaviour should not depend on the configuration used to prepare the current of OAM.
Since the usefulness of a large OHC is presumably the promise of being able to inject a current of OAM from one material into another then it isn't clear to us that it is a useful concept if this injection is not possible.
By arguing that an external electric field need not be applied in the ferromagnet for there to be long range transport of OAM when (near) orbital degeneracy occurs at the Fermi surface, this is implicitly recognized by Go {\it et al.} \cite{Go:prl23}.

None of the systems we have studied shows orbital diffusion on any significant length scale; indeed the decay is so rapid that it cannot be fit to an exponential form. 
Though the case had been made for short $l_{\rm of}$ in references \cite{Salemi:prm21, Belashchenko:prb23, Urazhdin:prb23}, the methods employed were not suitable for making quantitative estimates of $l_{\rm of}$ or taking into account various types of disorder. 
The extremely short length-scale we predict is at variance with the values extracted from experiment and disagrees with the idea that the orbital degrees of freedom act in a way physically similar to the spin \cite{Sala:prr22}, where the diffusive model has had considerable success in modeling experiments and describing quantum mechanical calculations alike \cite{Bass:jmmm16}.

On injecting an orbitally polarized current into Pt, which is known to have a high spin Hall angle, we saw that the orbital current is partially converted into a spin current within a few atomic layers, after which the spin current decays on a length scale determined by the spin flip diffusion length.
This potentially explains the long length scales extracted from experiment (Valet {\it et al.} argue that the electric field inhomogeneity at a boundary leads to an orbital-polarization profile with a distribution governed by the mean-free-path and that this is not evidence for currents of OAM \cite{Valet:prl25}). 
Through some mechanism (which likely depends on the experiment, the geometry of the sample and the materials involved), the orbital current might be converted into a spin current (or accumulation), so that the length scale measured is actually the spin flip diffusion length.
For the spin-orbit torque experiments, the theoretical prediction presented here indicates that any conversion layer between the orbital Hall source and the ferromagnetic layer, where the orbital current is converted into a spin current in order to exert the torque, need not be very thick at all. 
Additionally, the layer producing the orbital current through the orbital Hall effect need not be very thick either. 
The minimum thickness of this layer is dependent mostly on the Fuchs-Sondheimer suppression of the conductivity and its resulting current shunting, rather than the length scale on which orbital currents can be generated.

A positive result of the ``rediscovery'' \cite{Go:prl18} of the OHE \cite{Tanaka:prb08} has been to focus attention on light transition metals as the NM element in spintronics applications.
Although the OHE does not depend on SOC, the ability of an orbital polarization to exert torque on a magnetization does.
In the absence of evidence for bulk transport of orbital polarization, attention should focus on orbital accumulation at interfaces where the effective SOC can be large in spite of the received wisdom that a heavy metal was required to enhance the SHE because of the larger spin orbit coupling in the 5$d$ elements. 
This consideration neglects the importance of the narrower bandwidth of light transition metals and the effectiveness of interfaces in breaking the symmetry that suppresses the effect of SOC.
For example, the magnetocrystalline anisotropy energy (MAE) which results from SOC has a magnitude of only $\mu$eV/atom in cubic 3$d$ elemental metals (1.4$\mu$eV for bcc Fe and 2.7$\mu$eV for fcc Ni); the uniaxial symmetry of hcp Co already yields a MAE of $\sim 65 \mu$eV/atom \cite{Daalderop:prb90a}.
A monolayer of Co has a MAE of $\sim 1$meV/atom which results from the narrowing of the (already narrow) 3$d$ bandwidth by the reduced coordination number and depends strongly on the bandfilling \cite{Daalderop:prb94}. 
At an interface with a 4$d$ or 5$d$ metal, the MAE of these 3$d$ elements is as large as $\sim 1\,$meV/interface-atom and also depends on the bandfilling \cite{Daalderop:prb90b, Johnson:rpp96}.
Though the large SOC of the NM metal certainly plays a role, we note that the MAE resulting from the ``weak'' interface between the two 3$d$ elements Co and Ni has comparable size \cite{Daalderop:prl92, Johnson:rpp96}.

\section{Conclusion}
\label{sec:conc}

Using a first-principles implementation of Ando's scattering formalism for transport (which is equivalent in the linear response regime to the more conventional nonequilibrium Green's functions methods \cite{Khomyakov:prb05}), very large values of the OHC are predicted for light 3$d$ transition metals \cite{Rang:prb25} in qualitative agreement with results found using the Kubo formalism. 
The same scattering formalism finds that a current of orbital angular momentum injected into V, Cr, Cu or Pt does not propagate but decays on a length scale determined by electronic hopping and has at most a weak dependence on temperature-induced lattice disorder \cite{Rang:prb24}.  

\section*{Acknowledgements}
This work was sponsored by NWO Domain Science for the use of supercomputer facilities.


\begin{thebibliography}{70}
\bibitem [{foo()}]{footnote2}%
  \BibitemOpen
  \href@noop {} {}\bibinfo {note} {{Because of how currents are extracted from the scattering wave functions \cite{Wesselink:prb19}, $j(z)$ is not calculated in the leads. To visualise $j(z)$ in the left-hand lead ($\mathcal{L}$), the $\mathcal{L|S}$ interface is artificially shifted into the scattering region to $z$=$z_I$. The region $z<z_I$ where the magnetic field $B$ in \eqref{Eq:sZ} and \eqref{Eq:oZ} is nonzero and there is no thermal disorder is indicated in the figures by pink or blue shading. For simplicity, this effective lead will be referred to as the lead and the part of the scattering region $z<z_I$ will be referred to as the scattering region. }}\BibitemShut {NoStop}%
\end{thebibliography}

%

\end{document}